\documentclass[preprint,12pt]{elsarticle}
\usepackage[reviewcopy]{adndt}
\usepackage{longtable}

\usepackage{amsmath}

\usepackage{amssymb}

\begin{document}

\begin{frontmatter}

\journal{Atomic Data and Nuclear Data Tables}

%% Author, fill in article title here

\title{Effective oscillator strength distributions of spherically symmetric atoms
 for calculating polarizabilities and long-range atom-atom interactions}

%% Fill in author list here
%%  \author[one]{Author1\fnref{x}}
 \author[one,two]{Jun Jiang\corref{cor1}}
  \ead{phyjiang@yeah.net}

  \author[two]{J. Mitroy}
%  \ead{jim.mitroy@cdu.edu.au}
%  \author[Two]{C. Author}

  \author[two,three]{Yongjun Cheng }
  \ead{cyj83mail@gmail.com}

  \author[four]{M.~W.~J. Bromley }
  \ead{brom@physics.uq.edu.au}

  \cortext[cor1]{Corresponding author.}
%  \fntext[X]{First author footnote.}
%  \fntext[Y]{Second author footnote.}

\address[one]{Key Laboratory of Atomic and Molecular 
Physics and Functional Materials of Gansu Province,
College of Physics and Electronic Engineering, Northwest Normal
University, Lanzhou 730070, P. R. China}
\address[two]{School of Engineering, Charles Darwin University, Darwin, Northern Territory, 0909, Australia}
\address[three]{Academy of Fundamental and Interdisciplinary Science, Harbin Institute of Technology, Harbin 150080, P. R. China}
\address[four]{School of Mathematics and Physics,
The University of Queensland, Brisbane, Queensland 4075, Australia}
\date{22.10.2014} %please do not use \today, use actual date of version

\begin{abstract}

Effective oscillator strength distributions are systematically
generated and tabulated for the alkali atoms, the alkaline-earth atoms,
the alkaline-earth ions, the rare gases and some miscellaneous 
atoms.  These effective distributions are used to compute 
the dipole, quadrupole and octupole static polarizabilities,
and are then applied to the calculation of
the dynamic polarizabilities at imaginary frequencies. 
These polarizabilities can be used to determine
the long-range $C_6$, $C_8$ and $C_{10}$ atom-atom interactions
for the dimers formed from any of these atoms and ions, and we
present tables covering all of these combinations.

\end{abstract}
    
\end{frontmatter}

%%% Keywords and subject classification are not used in ADNDT
%%%\begin{keywords}
%%%Insert list of keywords here.
%%%\end{keywords}

%%% The table of contents should start a new page. This command will
%%% automatically pull all the unstarred \section, \subsection and
%%% \subsubsection titles into the Contents. Starred versions need to be
%%% done manually using
%%% \addcontentsline{toc}{[[sub]sub]section}{Section title}
%%% at the correct place. Examples are given below.

%%% The lists of data figures and data tables are created automatically
%%% by the \listofDfigures and \listofDtables commands.

\newpage

\tableofcontents
\listofDtables
\listofDfigures
\vskip5pc

%%%% Authors begin text of article here %%%

\section{Introduction}

The long-range interaction between two spherically 
symmetric atoms can be written in the general form
\cite{eisenschitz30a,margenau39a,fontana61a,mitroy03f}
\begin{equation}
V(R) \approx -\frac{C_6}{R^6} - \frac{C_8}{R^8} -\frac{C_{10}}{R^{10}} + \ldots \ ,
\end{equation}
where the $C_n$ parameters are the London/van der Waals 
dispersion coefficients. There are two complementary 
approaches to the computation of the dispersion 
coefficients. One approach uses oscillator strength 
sum-rules \cite{kumar85a,kumar85b}, while the second 
utilizes Casimir-Polder relations and uses the dynamic 
polarizabilities computed at imaginary energies
\cite{casimir48a,karplus64a}.  These approaches can 
be regarded as complementary to each other.

The key to the first approach is to generate an 
oscillator strength distribution that incorporates 
excitations to bound excited states and to the 
continuum states. In practice, the oscillator strength 
distributions are best termed `effective' oscillator 
strength distributions \cite{koide82a}. One might find
that the lowest few excited states are accurately represented 
by the distribution, however the higher bound states 
and continuum states are approximated with a set of discrete 
effective oscillator strengths and energies. The 
oscillator strength distributions can be derived from 
{\em ab-initio} structure calculations
\cite{maeder79a,yan96a,derevianko99a,mitroy03f},
experimental information such as refractive indices, 
atomic transition rates and photo-ionization cross 
sections \cite{kumar85a,kumar85b}, and sometimes both 
experimental and calculated oscillator strengths are 
used \cite{derevianko99a,porsev06a}.

The Casimir-Polder relation is reliant on being able to 
calculate the dipole and multipole dynamic polarizabilities 
at imaginary frequencies. One way to calculate a dynamic 
polarizability is to use oscillator strength sum-rules 
in conjunction with a previously determined oscillator
strength distribution. An alternate approach is to directly 
compute the dynamic polarizability as part of a structure 
calculation \cite{rahman90a,hattig96a,derevianko01a}.
The direct calculation of the dynamic polarizability is 
the preferred approach for structure calculations.

The present paper reports both effective oscillator 
strength distributions and dynamic polarizabilities 
for a number of spherically symmetric atoms and ions. 
The atoms presented are the noble gases, the alkali 
atoms and hydrogen, the singly-charged alkaline-earth 
ions and the alkaline earth atoms. The long-range 
atom-atom interaction coefficients $C_6$, $C_8$ 
and $C_{10}$ are also presented for any dimer formed 
from these atoms and ions. A previous tabulation of 
dynamic polarizabilities for many of these atoms does 
exist \cite{derevianko10a}. This previous tabulation 
only gave the dynamic dipole polarizabilities, while 
the present tabulation extends this to the quadrupole 
and octupole polarizabilities that are needed in the 
evaluation of $C_8$ and $C_{10}$.  The $C_8$ and $C_{10}$ 
dispersion coefficients are typically included in analysis 
of diatomic spectra aimed at characterizing the inter-atomic 
potential curve \cite{amiot02b,vanhaecke04a,meshkov14a}.  
At distances beyond the LeRoy radius \cite{leroy73a,ji95b} 
the inter-atomic interaction is reasonably well described 
by an interaction consisting exclusively of the dispersion 
interaction with the $C_6$, $C_8$ and $C_{10}$ terms 
\cite{mitroy03f,mitroy03g}.

\section{Definitions}

\subsection{Oscillator strength sum-rules}

All of the polarization parameters that are reported 
were computed from their respective oscillator strength 
sum-rules with the dipole, quadrupole and octupole 
oscillator strengths $f^{(\ell)}_{0i}$ from the ground 
state (with orbital and spin angular momentum equal zero) 
to the $i$th excited state defined \cite{yan96a,mitroy03f} 
as
\begin{equation}
f^{(\ell)}_{0i} =  \frac {2 |\langle \psi_0 \parallel r^{\ell}
{\bf C}^{\ell}({\bf \hat{r}} ) \parallel \psi_{i}\rangle|^2 \epsilon_{0i}}
{(2\ell+1) }  \ .
\label{fvaldef}
\end{equation}
In this expression ${\bf C}^{\ell}$ is the spherical 
tensor of rank $\ell$ while $\epsilon_{0i}$ is the 
excitation energy of the transition.  The sum-rule 
for the adiabatic multipole polarizability,
$\alpha^{(\ell)}$ \cite{dalgarno67a,mitroy03f} is
\begin{equation}
\alpha^{(\ell)} = \sum_{i} \frac {f^{(\ell)}_{0i} } {(\epsilon_{0i})^2}  \ .
\label{alphal1}
\end{equation}

One can also define other sum-rules
\cite{dalgarno63a,dalgarno66a,dalgarno67a}
such as
\begin{equation}
S^{(\ell)}(-k) = \sum_{i} \frac {f^{(\ell)}_{0i} } {(\epsilon_{0i})^{k}}  \ ,
\label{alphal2}
\end{equation}
where the $S^{(\ell)}(-2)$ are just the multipole 
polarizabilities.
The Thomas-Reiche-Kuhn (TRK) sum-rule says that $S^{(1)}(0)$ is equal
to the number of electrons in the atom \cite{kuhn25a,reiche25a,wang99a}.
There also is the atom-wall dispersion parameter, $C_3$ defined as \cite{kharchenko97a,johnson04b,mitroy03f}
\begin{equation}
C_3 = \frac{1}{8} \sum_{i} \frac {f^{(1)}_{0i} } {\epsilon_{0i}}
= \frac{1}{8} S^{(1)}(-1) \ .
\label{C3}
\end{equation}
These sum-rules are a generalized sum which implicitly 
includes a sum of excitations to ($E<0$) bound states and
an integration taking into account excitations to
($E>0$) continuum states. In the present work the sum-rule is
explicitly discretized, which is a consequence of diagonalization 
in a finite-sized box due to the finite range of the chosen
basis wave functions.

A similar application of the oscillator strengths is to
determine the standard atom-atom adiabatic dispersion 
parameters.  The dipole-dipole dispersion parameter between atoms 
A and B, $C_6$ is \cite{dalgarno67a,zhang12a}
\begin{equation}
C^{\rm AB}_6 = \frac{3}{2} \sum_{ij} \frac { f^{(1)}_{A,0i} f^{(1)}_{B,0j} }
    {\epsilon_{A,0i} \epsilon_{B,0j}(\epsilon_{B,0j}+\epsilon_{B,0i})} \ ,
\label{C6}
\end{equation}
the dipole-quadrupole dispersion parameter, 
$C_8$ is \cite{dalgarno67a,zhang12a}
\begin{equation}
C_8 = \frac{15}{4} \sum_{ij} \frac { f^{(1)}_{A,0i} f^{(2)}_{B,0j} }
    {\epsilon_{A,0i} \epsilon_{B,0j}(\epsilon_{A,0i}+\epsilon_{B,0j})}
    +  \frac{15}{4} \sum_{ij} \frac { f^{(1)}_{B,0i} f^{(2)}_{A,0j} }
    {\epsilon_{B,0i} \epsilon_{A,0j}(\epsilon_{B,0i}+\epsilon_{A,0j})} \ ,
\label{C8}
\end{equation}
and the dispersion parameter, 
$C_{10}$ is \cite{dalgarno67a,zhang12a}
\begin{eqnarray}
C_{10} &=& 7 \sum_{ij} \frac { f^{(1)}_{A,0i} f^{(3)}_{B,0j} }
    {\epsilon_{A,0i} \epsilon_{B,0j}(\epsilon_{A,0i}+\epsilon_{B,0j})}
    +  7 \sum_{ij} \frac { f^{(3)}_{A,0i} f^{(1)}_{B,0j} }
    {\epsilon_{A,0i} \epsilon_{B,0j}(\epsilon_{A,0i}+\epsilon_{B,0j})} \nonumber  \\
    &+&  \frac{35}{2} \sum_{ij} \frac { f^{(2)}_{A,0i} f^{(2)}_{B,0j} }
    {\epsilon_{A,0i} \epsilon_{B,0j}(\epsilon_{A,0i}+\epsilon_{B,0j})} \ .
\label{C10}
\end{eqnarray}

In a large dimension atomic structure calculation
the sum-rules defined in Eq.~\ref{alphal2} are calculated as:
\begin{equation}
S^{(\ell)}(-k) = \sum_{c=1}^{N_c} \frac {f^{(\ell)}_{c} } {\epsilon_{c}^{k}} 
+ \sum_{i=1}^{N} \frac {f^{(\ell)}_{i} } {\epsilon_{i}^{k}} 
\to \sum_{c=1}^{N_c} \frac {f^{(\ell)}_{c} } {\epsilon_{c}^{k}} 
+ \sum_{r=1}^{N_{r\prime}} \frac {f^{(\ell)}_{r} } {\epsilon_{r}^{k}} 
+ \sum_{r=N_{r\prime}}^{N_{r}} \frac {f^{(\ell)}_{r} } {\epsilon_{r}^{k}} 
+ \sum_{p} \frac {f^{(\ell)}_{p} } {\epsilon_{p}^{k}} \ .
\label{sk1}
\end{equation} 
Here, the first sum (with $f_c,\epsilon_{c}$) uses the core 
oscillator strength distribution for the number of core-electrons ($N_c$)
which is discussed later. The valence oscillator strength distribution
from a calculation of size $N$ can be logically broken up into the last
three terms. Both the second and third sums (with $f_r,\epsilon_{r}$) represent
the excitations to real (physical, $E<0$) bound states. The second sum
represents the contributions from the lowest, for example, $N_{r\prime}=4$ excited states
which typically account for more than 95$\%$ of the sum-rule for dipole 
excitations, while the excitations to more highly excited 
bound states are incorporated in the third sum. The fourth 
term (with $f_p,\epsilon_{p}$) contains the pseudo-oscillator strengths 
for excitations to ($E>0$) continuum states.

The multipole valence oscillator strength distributions typically
contain $N=15-6000$ terms for each multipole. Distributions with more
than $20$ terms for a given multipole are too unwieldy
for tabulation and easy computation. The solution to this problem
is to construct an effective oscillator strength distribution
($f_e,\epsilon_e$) \cite{langhoff74a,koide82a} 
in which a few effective oscillator strengths were adopted 
to represent the impact of the excitations to highly-excited 
bound states and continuum states, that is,
of the third and fourth sums in Eq.~\ref{sk1}.
In the present work, the sum-rule involving the effective oscillator
strength distributions is calculated as
\begin{equation}
S^{(\ell)}_{e}(-k) = \sum_{c=1}^{N_c} \frac {f^{(\ell)}_{c} } {\epsilon_{c}^{k}} 
+ \sum_{r=1}^{N_{r\prime}} \frac {f^{(\ell)}_{r} } {\epsilon_{r}^{k}} 
+ \sum_{e=1}^{N_e} \frac {f^{(\ell)}_{e} } {\epsilon_{e}^{k}} \ .
\label{sk2}
\end{equation}

The effective oscillator strength distribution 
($f_e,\epsilon_e$) is solved by setting the new 
sum-rules $S^{(\ell)}_{e}(-k)$ equal to the sum-rules 
$S^{(\ell)}(-k)$ calculated from large dimension 
atomic structure calculations. For example, a new 
distribution with two effective transitions 
($f_{e1},\varepsilon_{e1},f_{e2},\varepsilon_{e2}$)
is the solution of the non-linear equations constituted 
of four sum-rules: 
\begin{eqnarray}
S^{(\ell)}(0) & = &  \sum_{c=1}^{N_c} f_c + \sum_{r=1}^{N_{r\prime}}f_r + f_{e1} + f_{e2}  \\ 
S^{(\ell)}(-1)& = & \sum_{c=1}^{N_c} \frac{f_c}{\varepsilon_c} 
+ \sum_{r=1}^{N_{r\prime}} \frac{f_r}{\varepsilon_r} 
+ \frac{f_{e1}}{\varepsilon_{e1}} + \frac{f_{e2}}{\varepsilon_{e2}}   \\ 
S^{(\ell)}(-2)& = & \sum_{c=1}^{N_c} \frac{f_c}{\varepsilon_c^2} 
+ \sum_{r=1}^{N_{r\prime}} \frac{f_r}{\varepsilon_r^2} 
+ \frac{f_{e1}}{\varepsilon_{e1}^2} + \frac{f_{e2}}{\varepsilon_{e2}^2}   \\ 
S^{(\ell)}(-3)& = & \sum_{c=1}^{N_c} \frac{f_c}{\varepsilon_c^3} 
+ \sum_{r=1}^{N_{r\prime}} \frac{f_r}{\varepsilon_r^3} 
+ \frac{f_{e1}}{\varepsilon_{e1}^3} + \frac{f_{e2}}{\varepsilon_{e2}^3}  ,
\end{eqnarray}

\begin{figure}[th]
\caption{Schematic of the dipole transitions included in the sum-rules for lithium.
The $1s^2$ core electrons were assumed to only excite (via the dot-dashed
line) to a $1s 2\widetilde{p}$ pseudo-state with an energy $\Delta$.
The valence excitations are shown in two different ways as per
Eq.~\ref{sk1} and Eq.~\ref{sk2}:
(a) is the case of Eq.~\ref{sk1} where the primary transitions are shown
with solid connecting lines, whilst a large number of dashed-lines to
highly-excited bound states (up to $np$) are included as well as
pseudo-states (up to $n\bar{p}$) to describe the excitations to the continuum.
(b) is the case of Eq.~\ref{sk2} where the transitions to $2p_J$ and 
$3p_J$ states were still included while only three effective 
transitions (the dot-dashed lines to the $np'$ states) are adopted to represent
the contributions from excitations to the highly-excited bound states as well as the 
continuum states.}   
\label{fig1}
\vspace{0.1cm}
\includegraphics[width=8.4cm,angle=0]{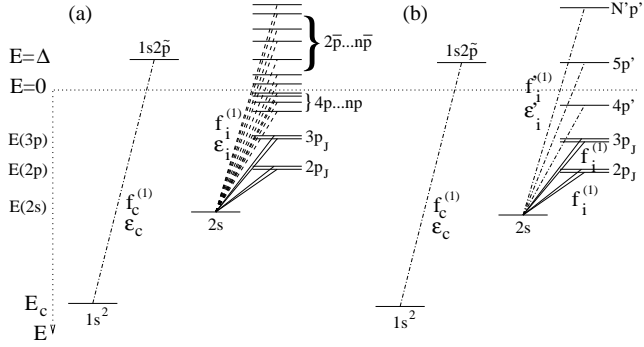}
\end{figure}

As an example, Fig.~\ref{fig1} depicts the dipole transitions 
included in the sum-rules for lithium. The core electrons 
were considered separately and assumed to only excite to 
a pseudo-state with an energy $\Delta$ for each multipole.
The resulting core oscillator distribution ($f_c,\epsilon_{c}$) 
is discussed later. The valence excitations were described 
in two different ways as in Eq.~\ref{sk1} and Eq.~\ref{sk2}. 

Table \ref{tabA} shows the numerical convergence of the dispersion 
coefficients of lithium dimer calculated using effective oscillator 
strength distributions with varying sizes. The oscillator strengths 
of the lowest four physical transitions for dipole and octupole 
and lowest six physical transitions for quadrupole were included 
in the calculations and the number of effective oscillator strengths 
varied from $N_e=0$ to $N_e=3$. The exact results were produced 
using Eq.~\ref{sk1}. A reasonable assessment is that each additional 
effective oscillator strength increases the accuracy by two orders 
of magnitude. Almost all of the effective oscillator strength 
distributions used in this paper have three effective oscillator 
strengths. This is sufficient to achieve an overall accuracy of 
at least five significant digits in the dispersion parameters. 
\renewcommand{\baselinestretch}{1.0}
\begin{table}[tbp]
\caption{Convergence of the $C_n$ dispersion parameters (in a.u.) 
for lithium dimer. The parameters are calculated using 
effective oscillator strength distributions with different sizes. 
$N_e$ gives the number of effective oscillator strengths that were
adopted. The `exact' results were calculated using Eq.~\ref{sk1}.
We thus adopt the $N_e=3$ set of effective oscillator strengths,
ie. for each multipole ($f_{e1}^{(\ell)},\varepsilon_{e1}^{(\ell)},
f_{e2}^{(\ell)},\varepsilon_{e2}^{(\ell)}$,
$f_{e3}^{(\ell)},\varepsilon_{e3}^{(\ell)}$),
which are given later in the paper.}
\label{tabA}
\begin{center}
\begin{tabular}{llll}
\hline
 $N_e$ & $C_6$ & $10^{-4} \times C_8$  & $10^{-6} \times C_{10}$ \\
 \hline\noalign{\vskip4pt}
$0$  & 1351.144   & 7.190653 & 3.899799 \\
$1$  & 1384.730   & 8.294263 & 7.294936 \\
$2$  & 1395.684   & 8.353312 & 7.381506 \\
$3$  & 1395.782   & 8.354554 & 7.382932 \\
exact& 1395.785   & 8.354584 & 7.382907 \\
\hline
\end{tabular}
\end{center}
\end{table}

Since there are three effective oscillator strengths adopted 
for almost all of the systems in the present work, six sum-rules 
were used to build a set of six non-linear equations and obtain
the effective oscillator strength distribution. The set of 
equations were solved with the Mathematica program 
which sometimes failed when the non-linear equations had a 
dimensionality of $6$. In such cases, the energy of the third 
effective oscillator strength was fixed manually and the 
dimensionality of the non-linear equations was reduced to be $5$. 
The energy of the third effective oscillator strength was 
adjusted manually until the non-linear equations gave correct 
solutions. 

\subsection{Casimir-Polder relations}

The dynamic polarizability at real frequencies is defined as
\begin{equation}
% \alpha^{(\ell)}(\omega) = \sum_{i} \frac {f^{(\ell)}_{0i} } {\omega^2-\epsilon_{0i}^2} \ ,
\alpha^{(\ell)}(\omega) = \sum_{i} \frac {f^{(\ell)}_{0i} } {\epsilon_{0i}^2 - \omega^2} \ ,
\label{alphal3}
\end{equation}
where a pole exists whenever the frequency is equal 
to a transition frequency.  At purely imaginary frequencies one has
\begin{equation}
\alpha^{(\ell)}(i\omega) = \sum_{i} \frac {f^{(\ell)}_{0i} } {\epsilon_{0i}^2 + \omega^2} \ ,
\label{alphal4}
\end{equation}
and there are no longer any poles.

The atom wall coefficient \cite{yan96a,kharchenko97a} can be written as
\begin{equation}
C_3 = \frac{1}{4 \pi} \int^{\infty}_{0} d\omega \ \alpha^{(1)}_{A}(i\omega) \: .
\end{equation}
The $C_6$ parameter is defined \cite{dalgarno66a,koide82a},
\begin{equation}
C_6 = \frac{3}{\pi} \int^{\infty}_{0} d\omega \ \alpha^{(1)}_{A}(i\omega)
\alpha^{(1)}_{B}(i\omega) \: ,
\end{equation}
while $C_8$ is \cite{dalgarno66a,koide82a}
\begin{eqnarray}
C_8 = \frac{15}{2\pi} \int^{\infty}_{0} d\omega \ \alpha^{(2)}_{A}(i\omega) \alpha^{(1)}_{B}(i\omega)
   +   \frac{15}{2\pi} \int^{\infty}_{0} d\omega \ \alpha^{(1)}_{A}(i\omega) \alpha^{(2)}_{B}(i\omega) \: ,
\end{eqnarray}
and $C_{10}$ \cite{dalgarno66a,koide82a} is
\begin{eqnarray}
C_{10} &=&  \frac{14}{\pi} \int^{\infty}_{0} d\omega \ \alpha^{(1)}_{A}(i\omega) \alpha^{(3)}_{B}(i\omega)
   +   \frac{14}{\pi} \int^{\infty}_{0} d\omega \ \alpha^{(3)}_{A}(i\omega) \alpha^{(1)}_{B}(i\omega) \nonumber \\
&+& \frac{35}{\pi} \int^{\infty}_{0} d\omega \ \alpha^{(2)}_{A}(i\omega) \alpha^{(2)}_{B}(i\omega) \: .
\end{eqnarray}

\subsection{Quadrature rules for Casimir-Polder integrations  }

The Casimir-Polder integrals are over the frequency 
range from zero to infinity. The integration mesh 
is constructed by mapping a Gauss-Legendre quadrature
grid defined over the $[0,1]$ interval to the $[0,\infty)$ 
interval using the transformation \cite{rossi11a}
\begin{equation}
\omega = \frac{au}{(1-u)}.
\end{equation}
Making this transformation leads to a set of weights 
and abscissa that constitute the quadrature rule. A 
uniform quadrature rule was adopted for the tabulation 
of the dynamic polarizabilities. A Gauss-Legendre
rule was adopted with the scaling parameter,
$a = 1.0$.

The accuracy of the quadrature rule was tested by 
performing the calculations on cesium. Values of 
$C_3$, $C_6$, $C_8$ and $C_{10}$ are calculated both 
directly from the oscillator strength sum-rules and 
from the Casimir-Polder integrals. Table \ref{tabB} 
shows that an integration rule with $40$ quadrature 
points is sufficient to give $C_6$ accurate to at 
least 8 significant digits. The values of $C_8$ 
and $C_{10}$ are even more accurate. Our least 
accurate dispersion parameter was $C_3$ which was 
precise to $5$ significant digits. Comparison with 
the quadrature rules of Derevianko \emph{et al.} 
(DPB) \cite{derevianko10a}, which used the 
transformation $\omega = 2 \tan(u\pi/2)$ 
\cite{derevianko10a} are also given. The DPB 
quadrature rules were not as precise as present 
quadrature rules, with a $50$ point rule giving 
only six digits accuracy for $C_6$.  
All of the calculations given in Table \ref{tabB} 
were done with the values computed to machine accuracy.
\begin{table}[tb]
\caption{Convergence of the atom-wall coefficient 
$C_{3}$ of cesium and dispersion coefficients $C_{n}$ 
for cesium dimer. The first `exact' row was computed
using the sums as per Eqs.~(\ref{C3}-\ref{C10})
including the full set of cesium states from the
large-basis calculation.
The other rows are all computed using Casimir-Polder 
integrals. The number of Gauss-Legendre integration 
points is $N$. The results labeled DPB-$N$ were 
computed using the $\tan$-based transformation formula 
\cite{derevianko10a}. Note that the present
calculations are all performed in full (double) precision
(ie. the tables of values given in Ref.~\cite{derevianko10a}
to six digits were not used).
All of the data is presented in a.u..}
\label{tabB}
\begin{center}
\begin{tabular}{lllll}
\hline
$N$    & \multicolumn{1}{c}{$C_3$} & \multicolumn{1}{c}{$C_6$} & \multicolumn{1}{c}{$C_8$} & \multicolumn{1}{c}{$C_{10}$} \\
\hline\noalign{\vskip4pt}
exact & 4.26007137597965    &    6732.75009597686    &    1003153.22405625    &    157922221.743034    \\
\hline
20    & 4.26019424843126    &    6732.66789142124    &    1003130.29211934    &    157920452.587935    \\
30    & 4.26002633795384    &    6732.74957771718    &    1003153.17110936    &    157922219.833385    \\
40    & 4.26006536827215    &    6732.75009436665    &    1003153.22396368    &    157922221.740791    \\ 
50    & 4.26008159786497    &    6732.75009594827    &    1003153.22405587    &    157922221.743012    \\ 
60    & 4.26007761977941    &    6732.75009597067    &    1003153.22405620    &    157922221.743030    \\ 
\hline
DPB-20 & 4.25538440902656    &    6696.52470846661    &    997816.300124684    &    157387242.062778    \\
DPB-30 & 4.26016857779331    &    6734.98988627677    &    1003351.22246868    &    157933531.124752    \\
DPB-40 & 4.26007587763244    &    6732.66175250520    &    1003148.54915587    &    157922091.763271    \\
DPB-50 & 4.26007949747275    &    6732.75267008174    &    1003153.29168060    &    157922221.708063    \\
%DPB-50 (6 digits) & 4.26007970136001    &    6732.75143321551    &    1003153.18972116    &    157922218.225063    \\
DPB-60 & 4.26007402950242    &    6732.75004218043    &    1003153.22411277    &    157922221.785811    \\
\hline
\end{tabular}
\end{center}
\end{table}

The dynamic polarizabilities tabulated later
are thus all given using a 40-point Gauss-Legendre rule 
with the scaling parameter set to $a = 1.0$.
For convenience example weights and abscissa for this rule
are given in Table \ref{tabC} to 7 significant figures,
although we do use the full machine precision
in our later calculations.  The discussion of the
effect that resulted from the use of the truncated values
from all of the tables is postponed towards the end
of the paper.
\begin{table}[tb]
\caption{Example Casimir-Polder integral grid locations ($\omega_i$) and weights
($w_i$) for 40-point integration. }
\label{tabC}
\begin{center}
\begin{tabular}{cccccc}
\hline
  $i$  &   $\omega_i$      &       $w_i$       & $i$  &   $\omega_i$     &   $w_i$   \\
  \hline
1  &   8.819222E$-$04  &   2.264628E$-$03  & 21  &   1.080673E$+$00  &   1.677693E$-$01 \\
2  &   4.658481E$-$03  &   5.298162E$-$03  & 22  &   1.262659E$+$00  &   1.972075E$-$01 \\
3  &   1.150079E$-$02  &   8.400470E$-$03  & 23  &   1.477386E$+$00  &   2.335615E$-$01 \\
4  &   2.149386E$-$02  &   1.160621E$-$02  & 24  &   1.732809E$+$00  &   2.790254E$-$01 \\
5  &   3.476135E$-$02  &   1.495651E$-$02  & 25  &   2.039486E$+$00  &   3.366805E$-$01 \\
6  &   5.147009E$-$02  &   1.849662E$-$02  & 26  &   2.411684E$+$00  &   4.109452E$-$01 \\
7  &   7.183462E$-$02  &   2.227705E$-$02  & 27  &   2.869039E$+$00  &   5.083035E$-$01 \\
8  &   9.612331E$-$02  &   2.635515E$-$02  & 28  &   3.439188E$+$00  &   6.385265E$-$01 \\
9  &   1.246661E$-$01  &   3.079702E$-$02  & 29  &   4.162008E$+$00  &   8.167931E$-$01 \\
10  &   1.578642E$-$01  &   3.567994E$-$02  & 30  &   5.096756E$+$00  &   1.067531E$+$00 \\
11  &   1.962032E$-$01  &   4.109535E$-$02  & 31  &   6.334558E$+$00  &   1.431716E$+$00 \\
12  &   2.402686E$-$01  &   4.715266E$-$02  & 32  &   8.021430E$+$00  &   1.981583E$+$00 \\
13  &   2.907663E$-$01  &   5.398426E$-$02  & 33  &   1.040330E$+$01  &   2.852384E$+$00 \\
14  &   3.485487E$-$01  &   6.175187E$-$02  & 34  &   1.392086E$+$01  &   4.317079E$+$00 \\
15  &   4.146481E$-$01  &   7.065504E$-$02  & 35  &   1.942876E$+$01  &   6.982043E$+$00 \\
16  &   4.903195E$-$01  &   8.094246E$-$02  & 36  &   2.876758E$+$01  &   1.237761E$+$01 \\
17  &   5.770978E$-$01  &   9.292713E$-$02  & 37  &   4.652492E$+$01  &   2.512244E$+$01 \\
18  &   6.768710E$-$01  &   1.070072E$-$01  & 38  &   8.695055E$+$01  &   6.351091E$+$01 \\
19  &   7.919797E$-$01  &   1.236948E$-$01  & 39  &   2.146622E$+$02  &   2.441386E$+$02 \\
20  &   9.253495E$-$01  &   1.436561E$-$01  & 40  &   1.133887E$+$03  &   2.911630E$+$03 \\
\hline
\end{tabular}
\end{center}
\end{table}

\section{Atomic Models}

The atomic models used to construct the oscillator 
strength distributions are discussed here. The static 
multipole polarizabilities and the atom-wall dispersion 
parameter, $C_3$, for all systems are given in Table \ref{tab1}.  
Comparisons with selected high accuracy calculations 
and/or experimental data are also incorporated in 
Table \ref{tab1}. These tabulations enable an initial 
assessment of the reliability of the atomic structure 
models for the dispersion coefficient calculations that 
are presented later.

\subsection{Hydrogen}

The hydrogen atom was diagonalized in a basis of $N=15$
Laguerre type orbitals for each angular momentum 
symmetry. Such a basis can give dispersion coefficients
to an accuracy of $13-14$ digits \cite{mitroy05a}. The 
resulting oscillator strength distribution does not 
include finite mass or relativistic effects. These 
two effects tend to cancel each other.

\subsection{Core oscillator strength distributions for multi-electron atoms}

The alkali and alkaline-earth atoms have both core 
and valence electrons. The oscillator strengths 
for the core were determined by using oscillator 
strength sum-rules as constraints \cite{rosenthal74a,mitroy03f},
i.e. the sum-rules computed using the core oscillator 
strength distribution must be equal to a previous
theoretical or experimental estimate of the mulitipole
polarizabilities of the core \cite{mitroy03c}. Especially, 
for dipole transitions one can use the TRK sum-rule as 
another constraint besides the core polatizability.

The initial estimate of the multipole oscillator 
strength for each shell is \cite{rosenthal74a,mitroy03f}:
\begin{equation}
f^{(\ell)}_i = \ell N_i \langle r^{2\ell-2}_i \rangle,
\label{osc}
\end{equation}
where $\langle r^{2\ell-2}_i \rangle$ expectation value 
is computed using the Hartree-Fock (HF) wave function 
for the core and $N_i$ is the number of electrons in each 
shell. Obviously, $f^{(\ell)}_i$ is equal to $N_i$ for 
dipole transitions. Then an energy shift $\Delta^{(\ell)}$
is applied to the Koopmans energy in order to make the 
core polarizabilities the same as the reference values
for each multipole. 

\begin{figure}[th]
\caption{Schematic of the dipole, quadrupole and octupole 
transitions to pseudo-states showing only the core excitations of Na. 
The $1s^22s^22p^6$ core electrons are assumed to only excite to
the pseudo-states $n\widetilde{p}$, $n\widetilde{d}$ and $n\widetilde{f}$
with energies $\Delta^{(1)}$, $\Delta^{(2)}$ and $\Delta^{(3)}$ 
via dipole, quadrupole and octupole transitions respectively. 
The oscillator strengths are $f^{(1)}_c$, $f^{(2)}_c$ and $f^{(3)}_c$ 
respectively. }   
\label{fig2}
\vspace{0.1cm}
\includegraphics[width=6cm,angle=0]{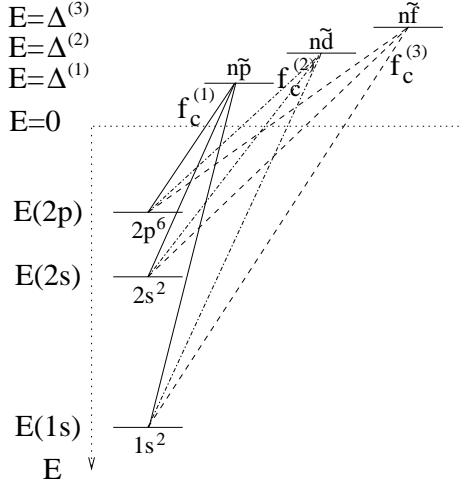}
\end{figure}

This technique can also be described as that all the core 
electrons are assumed to only excite to a pseudo-state with 
an energy of $\Delta^{(\ell)}$ and the related oscillator 
strength is set by Eq. \ref{osc}. As an example, Fig.~\ref{fig2} 
presentes how we derived the dipole, quadrupole and octupole 
core oscillator strengths of Na. The core electrons were 
assumed to only excite to pseudo-states with energies 
$\Delta^{(1)}$, $\Delta^{(2)}$ and $\Delta^{(3)}$ via 
dipole, quadrupole and octupole transitions respectively. 
The multipole oscillator strengths $f^{(1)}_c$, 
$f^{(2)}_c$ and $f^{(3)}_c$ were set by Eq.\ref{osc} and 
the energies $\Delta^{(1)}$, $\Delta^{(2)}$ and $\Delta^{(3)}$ 
were solved with constraints that the sum-rules are equal to 
the reference values of the dipole, quadrupole and octupole 
polarizabilities.  

\subsection{General comment for the alkali atoms and alkali-like ions}

The underlying structure models \cite{jiang13a} used for 
the Li$\to$Cs and Be$^+$$\to$Ba$^+$ sequences very closely 
resemble structure models used in previous non-relativistic 
calculations of these atoms \cite{mitroy03f}. Calculations 
are performed in a frozen core model, with the core taken 
from a Dirac-Fock calculation. A semi-empirical core polarization 
potential is used to incorporate the influence of core-valence 
correlations on the valence electron. The potential is adjusted 
so that the lowest energy of the state with a given ($\ell,j)$ 
is the same as the experimental energy. The wave functions
for the valence electrons are expanded in a basis of L-spinors 
\cite{grant00a,jiang13a}. The impact of core-polarization 
upon the multipole transition operators are included when 
the transition matrix elements are computed \cite{hameed68a}.
The reduction of the calculation to effectively a one-electron
problem makes it possible to eliminate basis set size as a 
significant source of error in the calculation.

For most of these atoms and ions, the binding energies of 
the lowest $ns$, $np_{J}$ and $nd_{J}$ systems were tuned 
to be very close to experiment, typically the differences 
with experimental energies were less than $10^{-6}$ Hartree. 
The energies of the next lowest excited states of each 
symmetry were not necessarily the same as experiment, but 
for the calculation of the dynamic polarizabilities these 
energies were adjusted manually to be the same as experiment.

\subsection{Lithium}

The present oscillator strength distributions are very 
similar to those used in a previous non-relativistic 
investigation of the dispersion coefficients of various 
low-lying states of the lithium dimer \cite{zhang07a}. 
It can be seen from Table \ref{tab1} that the present 
polarizabilities agree with Hylleraas calculations to 
an accuracy of about $0.1\%$.

\subsection{Sodium}

As in the case for lithium, a previous non-relativistic 
investigation of the sodium polarizabilities and dispersion 
coefficients has been made with a structure model very 
similar with the present calculation \cite{mitroy03f,mitroy05c}.  
The differences in polarizabilities from this earlier 
calculation do not exceed $0.1\%$. Polarizabilities from 
the present calculation are within $0.5\%$ of the polarizabilities 
from relativistic many-body perturbation theory calculations
(RMBPT) \cite{porsev03a,derevianko10a}.

\subsection{Potassium}

Details of the structure model can be found in previous 
works which investigated the tune-out wavelengths for 
potassium \cite{jiang13a,jiang13c}. Polarizabilities from
this calculations are within 0.5$\%$ of the polarizabilities 
from RMBPT calculations \cite{porsev03a,derevianko10a}.

\subsection{Rubidium}

The details of the model are very similar to a previous 
non-relativistic semi-empirical description of Rb \cite{mitroy03f}.
Numerical values from this earlier calculation (labeled CICP) 
are included in Table \ref{tab1} and are within 0.4$\%$ of 
the present polarizabilities. The polarizabilities from the 
RMBPT calculations \cite{derevianko10a} are all within 1$\%$ 
of the existing values.

\subsection{Cesium}

The calculations on cesium are completely new and follow a
methodology recently applied to other alkalis \cite{jiang13a,jiang13c}.
The present and RMBPT polarizabilities agree to better than 1$\%$.

\subsection{Copper and Silver}

The models used for Cu and Ag were non-relativistic.
The details of the model used to compute the oscillator 
strengths can be found in \cite{zhang08d}. The core
for these atoms contains a weakly bound $(n-1)d$ shell 
and the core polarizabilities are more than $10\%$ of 
the total polarizability. The oscillator strength 
distribution does not allow for physical excitations 
out of the $(n-1)d$ core. The notional precision of 
the polarizabilities for these two atoms is about $5-10\%$.

\subsection{Be$^+$}

The polarizabilities of Be$^{+}$ have previously been 
computed with a non-relativistic semi-empirical model \cite{tang09a}
and a Hylleraas calculation \cite{tang10a}. The present 
polarizabilities agree to better than $0.1\%$ with the 
Hylleraas calculation.

\subsection{Mg$^+$}

The details of the present model are very similar to 
a previous non-relativistic semi-empirical description 
of Mg$^+$ \cite{mitroy08b,mitroy09a}. The dipole 
polarizability of $34.99$ $a_0^3$ is very close to the 
experimental value of $35.04(3)$ $a_0^3$ \cite{snow08a,mitroy09a}. 
The higher order polarizabilities agree with RMBPT 
calculations at the $0.1\%$ level.

\subsection{Ca$^+$}

The details of the calculation are very similar to 
a previous non-relativistic semi-empirical description 
of Ca$^+$ \cite{mitroy08b}. All polarizabilities are 
in agreement with RMBPT calculations \cite{safronova11a}
to better than 1$\%$.

\subsection{Sr$^+$}

The model used for Sr$^+$ is relativistic. The details 
of the model are very similar to a previous non-relativistic 
semi-empirical description of Sr$^+$ \cite{mitroy08c}. All 
polarizabilities are in agreement with RMBPT calculations 
to better than 1.5$\%$.

\subsection{Ba$^+$}

The dipole and quadrupole polarizabilities are 2-3$\%$ 
smaller than RMBPT calculations \cite{iskrenova08a,safronova10d}.
There is a large difference of the octupole. The $\ell = 3$ 
orbitals are on the threshold of having an orbital collapse 
into an inner potential well \cite{lucatorto81a,cheng83a} 
and this makes the $nf$ orbitals, and polarizabilities,
relatively sensitive to very small changes in the interaction 
potential. So the 10$\%$ difference of the octupole with 
the RMBPT calculation was not surprising.

\subsection{Beryllium}

The Be transition matrix element list was obtained from 
a large dimension CI calculation with a semi-empirical 
polarization potential to represent core-valence correlations.
The semi-empirical potential included both one and two 
body polarization potentials \cite{mitroy03f,mitroy10d}.
For this calculation the energies of the lowest three dipole 
excited states were set to be the same as experiment \cite{cheng13b}.
These calculations are non-relativistic and hence do not 
explicitly include intercombination lines to the triplet 
states. Oscillator strengths for the transitions to the $^3P^o_1$  
states were added to the oscillator strength list by utilizing 
values from other atomic structure calculations \cite{cheng13b,fischer04a}. 
The intercombination transition has an insignificant affect
on the dipole polarizability except at frequencies very 
close to the transition frequency. The level of agreement 
with calculations \cite{komasa02a} that used explicitly 
correlated Gaussians (ECG) \cite{mitroy13a} 
is better than 0.2$\%$. The ECG calculations did not include 
relativistic or finite mass effects.

\subsection{Magnesium}

The structure calculations used to construct the oscillator 
strength distributions are described in Ref.~\cite{mitroy08a}.
The methodology is very similar to the approach adopted for 
beryllium.  The energies of the three lowest $^1P^o_1$ states
were set to experimental values \cite{nistasd500}. The level
of agreement with fully relativistic configuration interaction 
plus many body perturbation theory calculations (RCI+MBPT) 
\cite{porsev06a,derevianko10a} is better than 1$\%$ for the 
dipole and quadrupole polarizabilities with a 4$\%$ difference 
for the octupole polarizability. Oscillator strengths for the 
two lowest energy transitions to the $^3P^o_1$ states were 
added to the oscillator strength list by utilizing values from 
other atomic structure calculations \cite{cheng13b,fischer06a,MCHF09}.

\subsection{Calcium}

The structure calculations used to construct the oscillator 
strength distributions are described in Ref.~\cite{mitroy08g}.
The energies of the three lowest $^1P^o_1$ states were set 
to experiment \cite{nistasd500}. Oscillator strengths for 
the two lowest energy transitions to the $^3P^o_1$ states 
were added to the oscillator strength list by utilizing 
matrix elements from other atomic structure calculations
\cite{cheng13b,fischer03a}. The energies of the $^3P^o_1$ 
state were set to experiment. The polarizabilities agree 
with RCI+MBPT calculations \cite{porsev06a}
to an accuracy of better than  1.5$\%$ for the dipole 
and quadrupole transitions. The level of agreement is 
poorer for the octupole polarizabilities with a discrepancy
of 6$\%$.

\subsection{Strontium}

The structure calculations used to construct the oscillator 
strength distributions are described in Ref.~\cite{mitroy10b}.
The energies of the two lowest $^1P^o_1$ states were set 
to be the same as experiment  \cite{nistasd500}. The matrix 
element for the resonant transition was set to experiment \cite{mitroy10b}.
Oscillator strengths for the two lowest energy transitions 
to the $^3P^o_1$ states were added to the oscillator strength 
list by utilizing values from other atomic structure 
calculations \cite{cheng13b,safronova13b}. The energies of 
the triplet state were set to experiment \cite{nistasd500} .

\subsection{Barium}

The structure calculations for barium are new. These 
calculations follow the same non-relativistic technique 
used to determine wave functions and oscillator strengths
for strontium \cite{mitroy10b}.
The two electron binding energies of the lowest energy 
states were adjusted to be the same as experiment \cite{nistasd500}. 
The matrix element for the resonant $6s^2$ $^1S^e_0$ - $6s6p$ $^1P^o_1$ 
transition was taken from experiment \cite{bizzarri90a}. 
The oscillator strength for the intercombination line was 
added manually by utilizing a value from other atomic 
structure calculations \cite{cheng13b,dzuba06a}. The
energy of the triplet states were set to experiment.

\subsection{Helium($1s^2$)}

The helium wavefunctions were obtained using the 
configuration interaction (CI) method. The CI basis 
consisted of all possible configurations that could 
be formed from a single electron basis consisting of 
30 Laguerre type orbitals \cite{bromley01a} for each 
angular momentum with individual terms included up to 
$\ell = 5$. The parameter in the exponential term was 
set to 2.70. 
The polarizabilities agree to better than 0.05$\%$ with 
polarizabilities from Hylleraas calculations \cite{yan96a}. 
Omitted finite mass and relativistic effects are important 
at the 0.01$\%$ level of accuracy \cite{pachucki01a}.

\subsection{Neon, argon, krypton and xenon}

The dipole oscillator strength distributions for rare gas 
atoms are those of Kumar and Meath \cite{kumar85a,kumar85b}. 
These distributions were derived from discrete transition rates 
and photoionization cross sections that were constrained 
to be consistent with the experimental refractive index 
and the Thomas-Reiche-Kuhn sum-rules. The accuracy assigned
to the derived dispersion coefficients is typically of order 1$\%$.

The method to determine the quadrupole and octupole 
oscillator strengths for the rare gas atoms is similar 
to the method described in section 3.2 for core excitations. 
A detailed description of the treatment can be found 
in \cite{mitroy07d}. A noteable difference with the 
determination of the core oscillator strength distrbution 
is that one electron in the valence $np$ shell is treated
differently. This electron is assumed to excite to a 
pseudo-state with an energy $\Delta^{(\ell)}_b$ while
all of the other core electrons excite to a pseudo-state
with an energy $\Delta^{(\ell)}_a$. The multipole oscillator 
strengths are also set by Eq.\ref{osc} and the energies 
$\Delta^{(\ell)}_a$ and $\Delta^{(\ell)}_b$ are determined
by constraining the oscillator strengths sum-rules
to agree with sophisticated calculations of the multipole 
polarizabilities and the $C_8$ and $C_{10}$ coefficients 
for the homonuclear dimers
\cite{mitroy07d,thakkar92a,woon94a,nicklass95a,hattig96a}.

In the case of Kr and Xe, corrections were made to the 
reference polarizabilities to incorporate relativistic
effects. The accuracy of the quadrupole and octupole
oscillator strength distributions has been estimated 
elsewhere \cite{mitroy07d}. A reasonable 
estimate of the relative uncertainty in any sum-rule 
utilizing the quadrupole or octupole dynamic polarizabilities
would have 5$\%$ as the upper limit.

\subsection{He($1s2s \ ^3S^{e}$)}

The oscillator strength distribution used here was initially 
determined to compute the tune-out wavelengths of the helium 
metastable state \cite{mitroy13c}. The oscillator strength 
distribution was taken from a non-relativistic calculation 
computed within a frozen-core approximation in which the $1s$
core electron could excite to three states. This was labeled
as the CICP model \cite{mitroy13c}.

\subsection{Overview}

The accuracy of the polarizabilities is system dependent.
Results for the alkali atoms and alkali-like ions are 
generally the most accurate. The energies for the lowest 
dipole excited states are accurate to 10$^{-6}$ a.u.. The 
line strengths (i.e. the square of the reduced dipole matrix
elements) are accurate to 0.1$\%$ for Li and Be$^+$. The
accuracy degrades as the systems increase in size, and is
probably about 1-2$\%$ for Cs and Ba$^+$.

The accuracy achieved for the alkaline-earth atoms is 
comparable in quality to that achieved for the corresponding 
alkali atom in the same row of the periodic table. Although
the Hamiltonian is non-relativistic, the use of a tuned
polarization potential to some extent compensates for 
relativistic effects. These can be seen from the comparisons
between relativistic and non-relativistic polarizabilities 
of K using the present semi-empirical approach \cite{mitroy05b,jiang13a}.
The change in complexity going from one-electron atom to 
a two-electron atom hardly decreases the accuracy since 
the underlying configuration interaction calculations 
utilize very large basis with more than 150 orbitals.

The oscillator strength distributions for the alkali, and 
alkaline-earth ions will give reasonable descriptions of 
the actual dynamic dipole polarizabilities below the first 
two excitation thresholds. The dynamic polarizabilities 
for the alkaline-earth atoms will be reliable for energies
below the excitation threshold for the resonant
$ns^2$ $^{1}$D$^{e} \to nsnp \ ^1$P$^o$ transition.
The dynamic polarizabilities for copper and silver should
be reliable to the stated energies below the first excitation
threshold. The oscillator strength distributions of rare 
gases can be expected to become less reliable as the first 
excitation threshold is reached.

\section{Dynamic polarizability and dispersion coefficient tabulations}

\subsection{Real frequencies}

The effective oscillator strength distributions 
for dipole, quadrupole and octupole transitions
for all systems are given in Table \ref{tab2}. These 
distributions, by their design, give the same results
as the large calculations shown in Table \ref{tab1}.
They will also give reasonably accurate dynamic 
polarizabilities for real frequencies below the first
excitation thresholds, although these are not presented here.
Fortran programs that compute polarizabilities are
available on a web-site maintained by the authors \cite{vdwwww14a}.

\subsection{Imaginary frequencies}

The dynamic polarizabilities for dipole, quadrupole and
octupole transitions at imaginary energies are listed in 
Table \ref{tab3}.  They are computed at frequencies
based on a 40-point Gaussian rule that was used for
the consequent integrals presented in the paper.
Note that the grids and integration weights used for the
frequency integrations are given in Table \ref{tabC}
to 7 significant figures, although the calculations
in Table \ref{tab3} were performed at machine precision.
%Table \ref{tab3} gives the
%dynamic dipole polarizability at imaginary energies for all system.

\subsection{Dispersion coefficients}

Table \ref{tab4} gives the $C_6$, $C_8$ and $C_{10}$ 
dispersion coefficients for all combinations 
of the atoms listed in the article. This table was 
generated using the effective oscillator strength
distributions (using full machine precision) using
$40$ point integrations.
The $C_6$ parameters in the tabulation lie very close to those 
of the Derevianko {\em et al.} tabulation \cite{derevianko10a} 
for systems involving the same atoms. The largest discrepancy 
for a homo-nuclear dimer occurs for barium and was 4$\%$.
The relative accuracy of the other dispersion coefficients can be estimated
by reference to the text above and via the references where the
initial versions of our oscillator strengths were first computed.

The long-range dispersion coefficients are also given for
the ion-atom dimers, such as Mg$^+$-He.  The ion-atom interaction,
however, also contains a (stronger) long-range polarization interaction.
This has the form $V \sim -\frac12 \alpha_d/R^4$ where $\alpha_d$ is the
static polarizability of the neutral atom \cite{mitroy08b}.

Fortran programs that compute dispersion coefficients are
available on a web-site maintained by the authors \cite{vdwwww14a}.

\subsection{Impact of finite precision tabulations}

The final issue to consider here is the impact of using the
tabulated values which are presented with less precision
than were used in the actual calculations.
As an example, reconsider the case of the dispersion coefficients
of cesium as shown in Table \ref{tabB}.
With the $40$-point integration rule the full machine precision
calculations gave
$C_3=4.26006536827215$, $C_6=6732.75009436665$, $C_8=1003153.22396368$, $C_{10}=157922221.740791$.
If one uses the effective oscillator strengths as tabulated
in Table \ref{tabB} to compute the polarizabilities,
at the frequencies tabulated in Table \ref{tabC}, and
then perform the integrations, then the result is
$C_3=4.2600653298881$, $C_6=6732.74980359221$,
$C_8=1003153.18636727$, $C_{10}=157922216.822495$.
Thus the finite precision is not an issue since the
noise generated by the truncation is much lower than the
difference between the present dispersion coefficients
and other accurate theoretical calculations.

\section{Conclusion}

Effective oscillator strength distributions and dynamic 
dipole polarizabilities at imaginary frequencies have been 
given for eighteen atoms and five alkali-like ions. All the 
atoms and ions are in a spherically symmetric state. 
Providing the effective oscillator strength distributions
means that the dynamic polarizabilities are not restricted 
to the frequencies tabulated here.

The present work should be regarded as the first step in 
an ongoing effort that aims to provide a data repository 
that will facilitate the determination of long-range 
interactions for more pairs of atomic and molecular systems \cite{vdwwww14a}.

\ack This research was supported by the Australian Research Council
Discovery Project DP-1092620. The work of JJ was supported 
by the National Natural Science Foundation of China under Grant
No. 11147018 and the basic scientific research foundation of institution 
of higher learning of Gansu Province. The work of YC was supported by the National Natural 
Science Foundation of China under Grant No. 11304063. MWJB was 
supported by the Australian Research Council Future Fellowship FT100100905,
and thanks Julia M. Rossi for her work exploring the convergence
of the Gaussian integration scheme that was used here.

This work is dedicated to Professor James (Jim) Mitroy,
our colleague and mentor, who unexpectedly passed away shortly
after the initial submission of this manuscript.

%%  All sections inside the appendix environment will be appendixes
%%  Subsections function normally in appendixes.
%
%\begin{appendix}
%
%\def\thesection{} % To get the appendix heading correct
%
%\noteinproof
%A note added in proof, if there is one, should be the final text before
%the references.

\bibliographystyle{model1a-num-names}

% \bibliography{atombrommit,adndtbst}
% \bibliography{atombrommit}

‎

%\begin{thebibliography}{99}
%\bibitem{87Ram}
%        S. Raman, C. H. Malarkey, W. T. Milner, C. W. Nestor, Jr.,
%        P. H. Stelson, Atomic Data and Nuclear Data Tables
%        36 (1987) 1.
%\end{thebibliography}

\clearpage

\TableExplanation

\bigskip
\renewcommand{\arraystretch}{1.0}

\section*{Table 1.\label{tabDE} Multipole polarizabilities and atom-wall
dispersion parameter for all atoms and ions.}

The first row for any system is computed from the oscillator strength distributions
used in this work. The other rows for each system give polarizabilities taken from other sources.

\begin{tabular}{@{}p{1in}p{6in}@{}}
$C_3$           & atom-wall dispersion parameter in a.u.\\
$\alpha_1$      & static dipole polarizability in a.u.\\
$\alpha_2$      & static quadrupole polarizability in a.u.\\
$\alpha_3$      & static octupole  polarizability in a.u.\\
Hyl             & Hylleraas or correlated Slater basis  \\
CICP            & Configuration interaction with semi-empirical core-valence interaction  \\
RMBPT           & relativistic many-body perturbation theory \\
RCI+MBPT        & relativistic configuration interaction with many-body perturbation theory to incorporate core-valence correlations.  An asterisk is used to denote polarizabilities where the calculation matrix element for the resonance transition has been replaced by an experimental value.   \\
ECG             & explicitly correlated Gaussian basis \\
RCISD           & relativistic configuration interaction calculations with single and double excitations. \\
RCCSD           & relativistic coupled-cluster calculations and the contributions of singly and doubly excited states were considered. \\
Expt.           & experimental value \\
\end{tabular}
\label{tableII}

\bigskip
\renewcommand{\arraystretch}{1.0}

\section*{Table 2.\label{tabPF} The dipole, quadrupole, and octupole effective
oscillator strength distributions. }

The oscillator strengths are in the $f$ column and excitation energies
(in a.u.) are given in the $\varepsilon$ column. Effective oscillator
strengths for alkali and alkaline-earth systems are given in the
following order. Firstly the oscillator strengths for core
excitations are given. These can be identified easily since the
oscillator strength is usually a whole number, such as 2.0.  Next,
the oscillator strengths for physical transitions are listed.
Finally, the effective oscillator strengths that account for
transitions to the higher excited states and continuum are listed.

Different superscripts $c$, $r$, $p$ and $e$ are used to identify 
each transition corresponding to core excitations ($c$, computed as 
per \cite{mitroy03f}), real physical excitations ($r$), excitations 
to pseudo-states ($p$) and effective excitations ($e$). The calculations 
of the core for He($1s2s \ ^3S^{e}$) adopted a different method 
that can be found in \cite{mitroy13c} and thus they are labeled by 
the superscript $d$. The oscillator strength distributions of the 
rare gas atoms are computed as per \cite{kumar85a,kumar85b,mitroy07d} 
and they are labeled by the superscript $g$. 

\begin{tabular}{@{}p{1in}p{6in}@{}}
$\ell$              & the order of multipole \\
$\varepsilon$  & excitation energy (in a.u.) \\
$f$            & oscillator strength \\
Superscript $c$  & core excitations \\
Superscript $r$  & real physical excitations \\
Superscript $p$  & excitaions to pseudo-states \\
Superscript $e$  & effective excitations \\
Superscript $d$  & core excitations for He($1s2s \ ^3S^{e}$) \\
Superscript $g$  & rare gas atoms \\
\end{tabular}
%\label{tableII}

\section*{Table 3.\label{tabDP} The dipole, quadrupole, and octupole 
dynamic polarizabilities of all atoms and ions (in a.u.).}

The dynamic dipole, quadrupole, and octupole polarizabilities (in $a_0^3$) 
at a set of photon energies that constitute a quadrature rule for the 
Casimir-Polder integral. The energies and integration weights are given 
in Table \ref{tabC}.

\section*{Table 4.\label{tabCN} The dispersion coefficients, 
$C_6$, $C_8$ and $C_{10}$ (in a.u.). }
The dispersion coefficients, $C_6$, $C_8$ and $C_{10}$ (in a.u.)
for all possible pairs of atoms and ions. The He I(T) represents 
the He $1s2s \ ^3S^{e}$.

\datatables
\setlength{\tabcolsep}{0.5\tabcolsep}
\renewcommand{\arraystretch}{1.0}

\footnotesize % we need to squeeze the font size a lot!
\begin{longtable}{@{\extracolsep\fill}lllll@{}}
\caption{Multipole static polarizabilities and atom-wall dispersion 
parameter for all atoms and ions. See Page\ \pageref{tabDE} for explanation of tables.}
\label{tab1}
System  & $C_{3}$ & $\alpha_1$  & $\alpha_2$ & $\alpha_3$ \\
 \hline
H  &  0.250   &  4.50       & 15.0       & 131.25      \\
He &  0.188132   & 1.38333   & 2.44596 & 10.6262 \\
   &  0.188131 Hyl. \cite{yan96a}  & 1.38319  Hyl. \cite{yan96a} & 2.44508 Hyl. \cite{yan96a} & 10.6203 Hyl. \cite{yan96a} \\
He($1s2s \ ^3S^e$) & 1.900      & 316.0  & $2.708 \times 10^3$  & $8.837 \times 10^4$                \\
                   & 1.90092 Hyl. \cite{yan98a}  & 315.63 Hyl. \cite{yan98a}  &  $2.7079 \times 10^3$ Hyl. \cite{yan98a} & $8.8377 \times 10^4$ Hyl. \cite{yan98a}  \\
Li & 1.524                   & 164.3                  & $1.424 \times 10^3$                 & $3.968 \times 10^4$                 \\
   & 1.518001 Hyl. \cite{yan95b} & 164.111 Hyl. \cite{yan96a}  & $1.42327 \times 10^3$ Hyl. \cite{yan96a} & $3.96505 \times 10^4$ Hyl. \cite{yan96a} \\
\hline
\end{longtable}

%%% Please start a new page by uncommenting the next

\newpage
%% If the table is to span over the whole text width, we set:

%\setlength{\LTleft}{0pt}
%\setlength{\LTright}{0pt} 

% To avoid ``Overfull \hboxes...'' decrease the intercolumn spacing:

\setlength{\tabcolsep}{0.5\tabcolsep}
\renewcommand{\arraystretch}{1.0}
\footnotesize % we need to squeeze the font size a lot!

\begin{longtable}{@{\extracolsep\fill}ccccccccccc@{}}
\caption{The dipole, quadrupole and octupole effective oscillator strength distributions. See Page\ \pageref{tabPF} for explanation of tables.}
\label{tab2}
  $\ell$ &   $\epsilon$ & $f$ &&  $\ell$ & $\epsilon$ & $f$ &&  $\ell$ & $\epsilon$ & $f$\\
\hline
\endfirsthead
\caption[]{(continued)}
  $\ell$ &   $\epsilon$ & $f$ &&  $\ell$ & $\epsilon$ & $f$ &&  $\ell$ & $\epsilon$ & $f$\\
\hline
\endhead
\multicolumn{11}{c}{H I} \\
     1$^r$ &  3.7500E$-$01  &  4.1620E$-$01  && 2$^r$ &  4.4445E$-$01  &  5.3412E$-$01  &&  3$^r$ &  4.6892E$-$01  &  8.0763E$-$01    \\
     1$^r$ &  4.4446E$-$01  &  7.9290E$-$02  && 2$^r$ &  4.6959E$-$01  &  2.8671E$-$01  &&  3$^r$ &  4.8494E$-$01  &  1.3974E$+$00    \\
     1$^r$ &  4.7102E$-$01  &  4.1698E$-$02  && 2$^r$ &  4.9081E$-$01  &  3.9301E$-$01  &&  3$^p$ &  5.0836E$-$01  &  3.0489E$+$00    \\
     1$^r$ &  4.9895E$-$01  &  5.4677E$-$02  && 2$^p$ &  5.2305E$-$01  &  5.6031E$-$01  &&  3$^p$ &  5.4308E$-$01  &  5.4465E$+$00    \\
     1$^p$ &  5.4195E$-$01  &  6.5297E$-$02  && 2$^p$ &  5.6966E$-$01  &  6.9724E$-$01  &&  3$^p$ &  5.9155E$-$01  &  8.1216E$+$00    \\
     1$^p$ &  6.0433E$-$01  &  6.9491E$-$02  && 2$^p$ &  6.3474E$-$01  &  7.7516E$-$01  &&  3$^p$ &  6.5780E$-$01  &  1.0266E$+$01    \\
     1$^p$ &  6.9307E$-$01  &  6.7969E$-$02  && 2$^p$ &  7.2510E$-$01  &  7.7342E$-$01  &&  3$^p$ &  7.4828E$-$01  &  1.1071E$+$01    \\
\end{longtable}

%\newpage
%\include{table3}

\newpage
%% If the table is to span over the whole text width, we set:

%\setlength{\LTleft}{0pt}
%\setlength{\LTright}{0pt} 

% To avoid ``Overfull \hboxes...'' decrease the intercolumn spacing:

\setlength{\tabcolsep}{0.5\tabcolsep}
\renewcommand{\arraystretch}{1.0}
\footnotesize % we need to squeeze the font size a lot!

\begin{longtable}{@{\extracolsep\fill}ccccccccc@{}}
\caption{The dipole, quadrupole and octupole dynamic polarizabilities (in a.u.) for all atoms and ions. See Page\ \pageref{tabDP} for explanation of tables.}
\label{tab3}
 $i$ &  $\alpha^{(1)}_i$ &  $\alpha^{(2)}_i$ &   $\alpha^{(3)}_i$ &&  $i$ & $\alpha^{(1)}_i$ &  $\alpha^{(2)}_i$ &   $\alpha^{(3)}_i$  \\
\hline
\endfirsthead
\caption[]{(continued)}
 $i$ & $\alpha^{(1)}_i$ &  $\alpha^{(2)}_i$ &   $\alpha^{(3)}_i$ &&  $i$ & $\alpha^{(1)}_i$ &  $\alpha^{(2)}_i$ &   $\alpha^{(3)}_i$  \\
\hline
\endhead
\multicolumn{4}{c}{H I} && \multicolumn{4}{c}{Li I}  \\
    1  &         4.5000E$+$00 &      1.5000E$+$01 &      1.3125E$+$02 &&    1  &         1.6424E$+$02 &      1.4240E$+$03 &      3.9676E$+$04  \\
    2  &         4.4994E$+$00 &      1.4999E$+$01 &      1.3124E$+$02 &&    2  &         1.6351E$+$02 &      1.4227E$+$03 &      3.9654E$+$04  \\  
    3  &         4.4965E$+$00 &      1.4993E$+$01 &      1.3121E$+$02 &&    3  &         1.5974E$+$02 &      1.4158E$+$03 &      3.9535E$+$04  \\
    4  &         4.4878E$+$00 &      1.4977E$+$01 &      1.3110E$+$02 &&    4  &         1.4948E$+$02 &      1.3956E$+$03 &      3.9184E$+$04  \\
    5  &         4.4681E$+$00 &      1.4940E$+$01 &      1.3085E$+$02 &&    5  &         1.3055E$+$02 &      1.3520E$+$03 &      3.8415E$+$04  \\
    6  &         4.4308E$+$00 &      1.4869E$+$01 &      1.3038E$+$02 &&    6  &         1.0499E$+$02 &      1.2754E$+$03 &      3.7017E$+$04  \\
    7  &         4.3673E$+$00 &      1.4747E$+$01 &      1.2956E$+$02 &&    7  &         7.8444E$+$01 &      1.1613E$+$03 &      3.4826E$+$04  \\
\end{longtable}

\newpage
%% If the table is to span over the whole text width, we set:

%\setlength{\LTleft}{0pt}
%\setlength{\LTright}{0pt} 

% To avoid ``Overfull \hboxes...'' decrease the intercolumn spacing:

\setlength{\tabcolsep}{0.5\tabcolsep}
\renewcommand{\arraystretch}{1.0}
\footnotesize % we need to squeeze the font size a lot!

\begin{longtable}{@{\extracolsep\fill}lllr@{}}
\caption{Dispersion coefficients of $C_6$, $C_8$ and $C_{10}$. See Page\ \pageref{tabCN} for explanation of tables.}
\label{tab4}
    &  $C_6$ & $C_8$ & $C_{10}$\\
\hline
\endfirsthead
\caption[]{(continued)}
    &  $C_6$ & $C_8$ & $C_{10}$\\
\hline
\endhead
  H I   - H I     &    6.4990E+00 &   1.2440E+02 &   3.2858E+03 \\
  H I   - Li I    &    6.6591E+01 &   3.2824E+03 &   2.2318E+05 \\
  H I   - Na I    &    7.4183E+01 &   4.0098E+03 &   2.9180E+05 \\
  H I   - K I     &    1.1203E+02 &   7.9650E+03 &   7.3391E+05 \\
  H I   - Rb I    &    1.2496E+02 &   9.5418E+03 &   9.3170E+05 \\
  H I   - Cs I    &    1.4772E+02 &   1.2956E+04 &   1.4037E+06 \\
  H I   - Cu I    &    3.8328E+01 &   1.3323E+03 &   5.5191E+04 \\
  H I   - Ag I    &    4.5544E+01 &   1.6569E+03 &   6.9788E+04 \\
\end{longtable}

\end{document}